\documentclass[10pt,twocolumn,prl,aps,amssymb,amsmath,tightenlines]{revtex4} 

\usepackage[caption = false]{subfig}
\usepackage{dsfont}
\usepackage{graphicx}
\usepackage{amssymb,amsfonts,amsthm}
\usepackage{color}
\usepackage{bm,float}

\newcommand{\half}{\mbox{$\textstyle \frac{1}{2}$}}

\newcommand{\re}{\mbox{$\rm e$}}
\newcommand{\ri}{\mbox{$\rm i$}}
\newcommand{\rd}{\mbox{$\rm d$}}

\begin{document}

\title{Phase-space measurements, decoherence and classicality}

\author {\textsc{Dorje C.~Brody$^{1,2}$, Eva-Maria Graefe$^{3}$, and 
Rishindra Melanathuru$^{3}$}} 

\affiliation{$^{1}$School of Mathematics and Physics, University of Surrey, 
Guildford GU2 7XH, UK \\ 
$^{2}$Institute of Industrial Science, The University of Tokyo, Tokyo 153-0041, Japan \\
$^{3}$Department of Mathematics, Imperial College London, London SW7 2BZ, UK 
}

\begin{abstract} 
The emergence of classical behaviour in quantum theory is often ascribed to the 
interaction of a quantum system with its environment, which can be interpreted 
as environmental monitoring of the system. As a result, off-diagonal elements of 
the density matrix of the system are damped in the basis of a preferred observable, 
often taken to be the position, leading to the phenomenon of decoherence. This 
effect can be modelled dynamically in terms of a Lindblad equation driven by the 
position operator. Here the question of decoherence resulting from a monitoring of 
position \textit{and} momentum, i.e. a phase-space measurement, by the environment 
is addressed. There is no standard quantum observable corresponding to the detection 
of phase-space points, which is forbidden by Heisenberg's uncertainty principle. This 
issue is addressed by use of a coherent-state-based positive operator-valued measure 
(POVM) for modelling phase-space monitoring by the environment. In this scheme, 
decoherence in phase space implies the diagonalisation of the density matrix in both 
position and momentum representations. This is shown to be linked to a Lindblad 
dynamics where position and momentum appear as two independent Lindblad 
operators. 
\end{abstract}

\maketitle

The notion of decoherence plays an important role in modern quantum theory for 
characterising the emergence of classicality \cite{Zurek1991,Zurek,JZ,MS,Zurek22}. 
Heuristically, the phenomena can be described as the decay of the off-diagonal 
elements of the density matrix in the basis of a preferred observable selected by 
the system-environment interaction. The dynamics of the state can be modelled 
dynamically in terms of 
a Lindblad equation generated by that observable. 
For example if the initial state of the system is a 
superposition of a particle being ``here \textit{and} there'' and if the preferred observable 
is the position of the particle, then after decoherence we have a mixed state that 
represents the particle being ``here \textit{or} there'' \cite{Halliwell}. 
In the literature there has been an emphasis on this position-space decoherence, 
physically motivated by the scattering of a quantum particle by air molecules, through which 
the position of the particle is in effect being monitored \cite{Schlosshauer 2019}. 

\begin{figure}[t!]
      \centering
\hspace{-0.2cm} \includegraphics[width=0.16\textwidth]{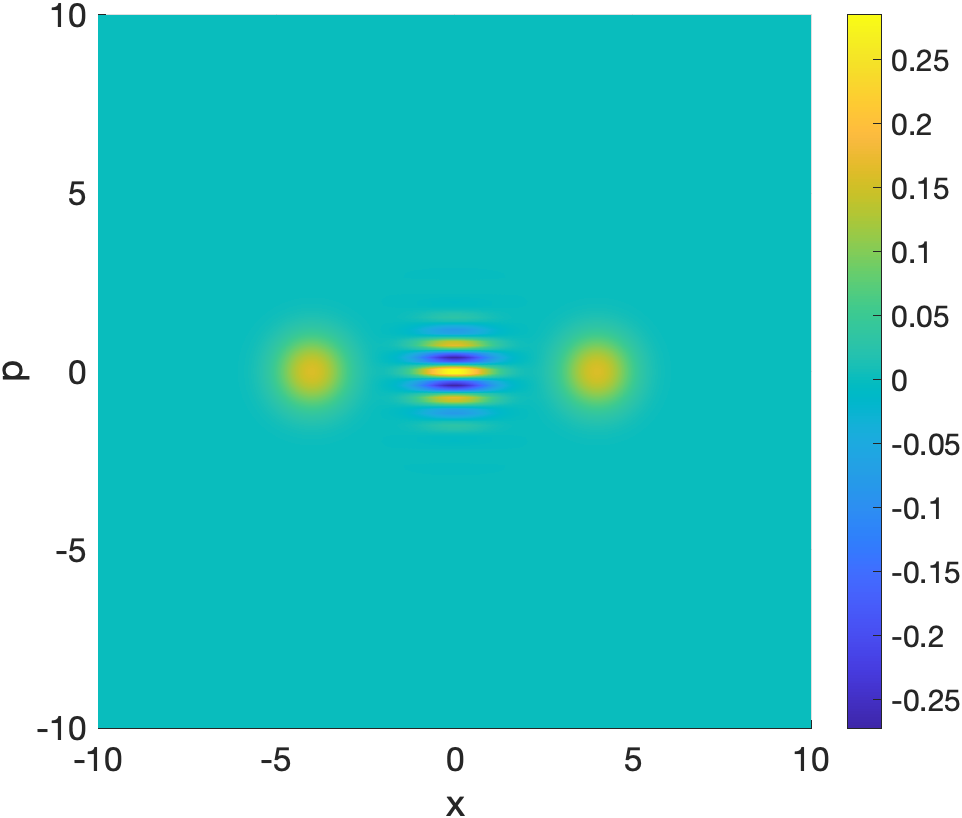} 
 \hspace{-0.2cm} \includegraphics[width=0.16\textwidth]{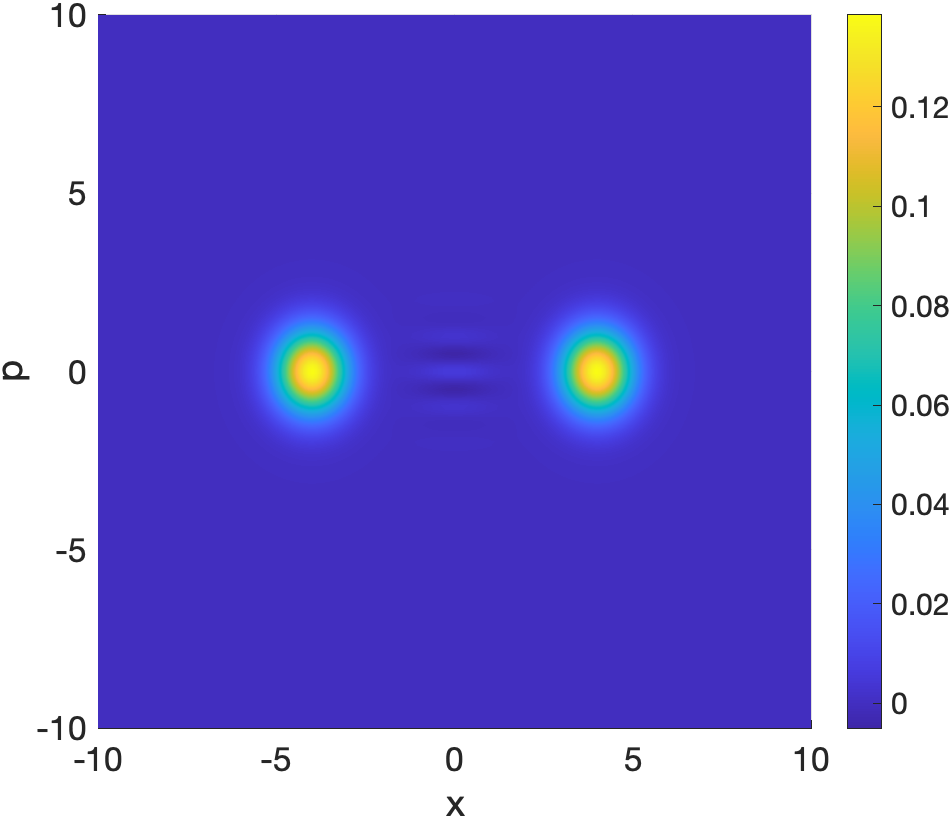}
  \hspace{-0.2cm} \includegraphics[width=0.16\textwidth]{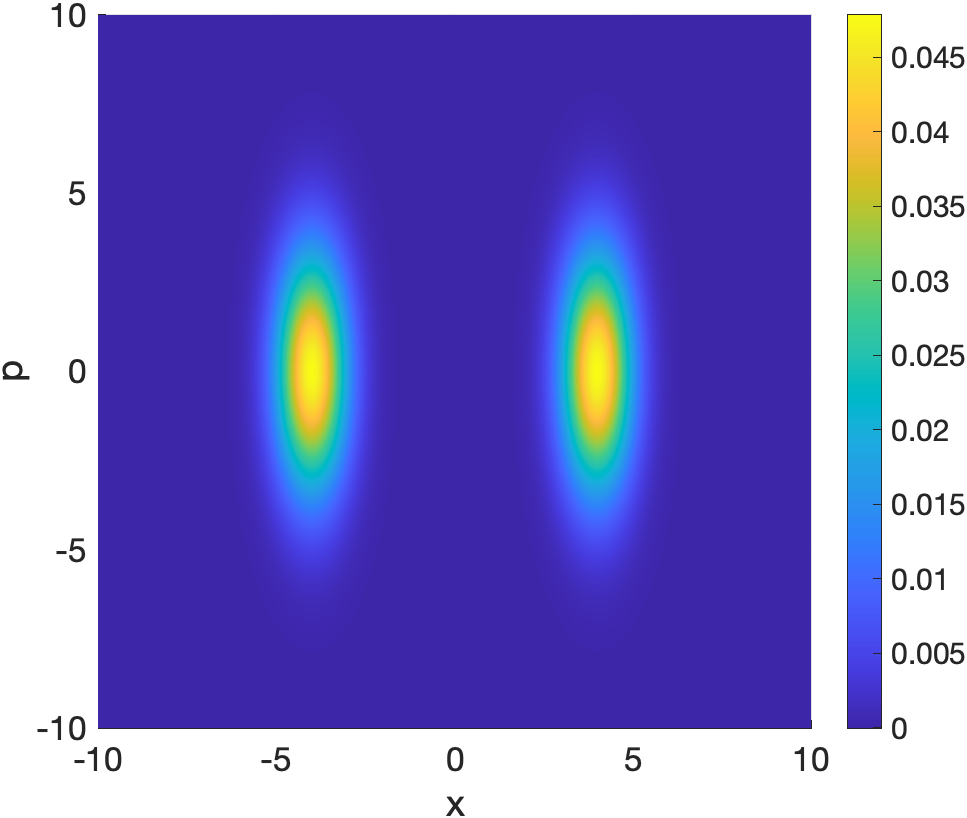}\\
  \hspace{-0.2cm} \includegraphics[width=0.16\textwidth]{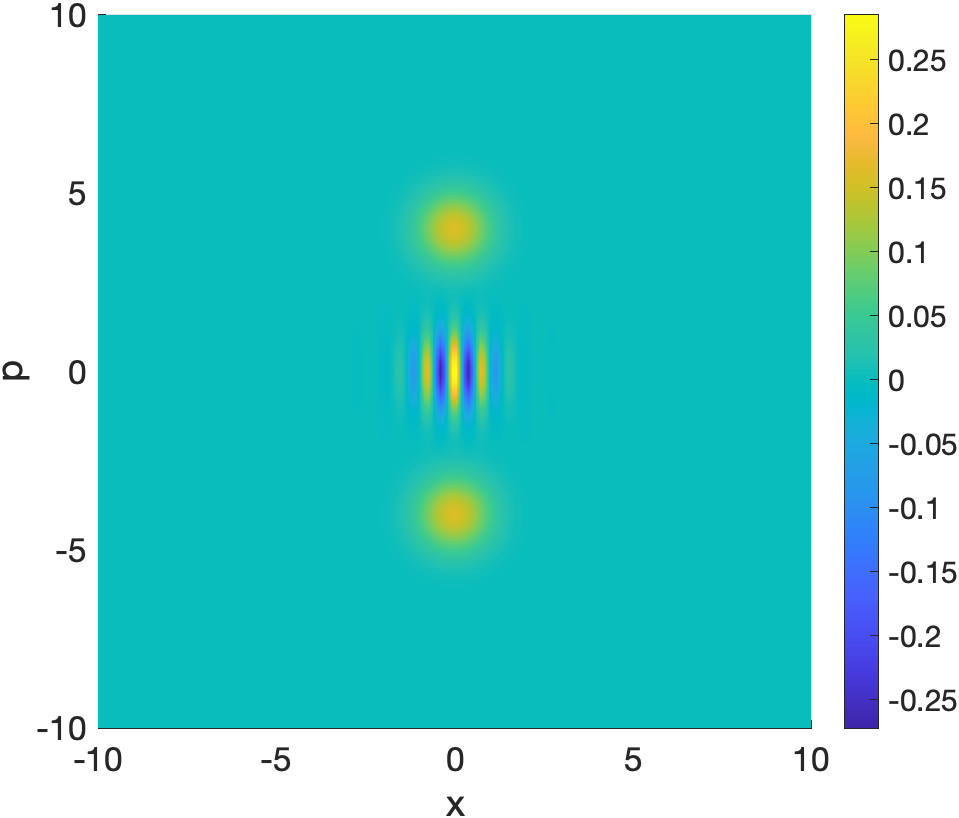} 
 \hspace{-0.2cm} \includegraphics[width=0.16\textwidth]{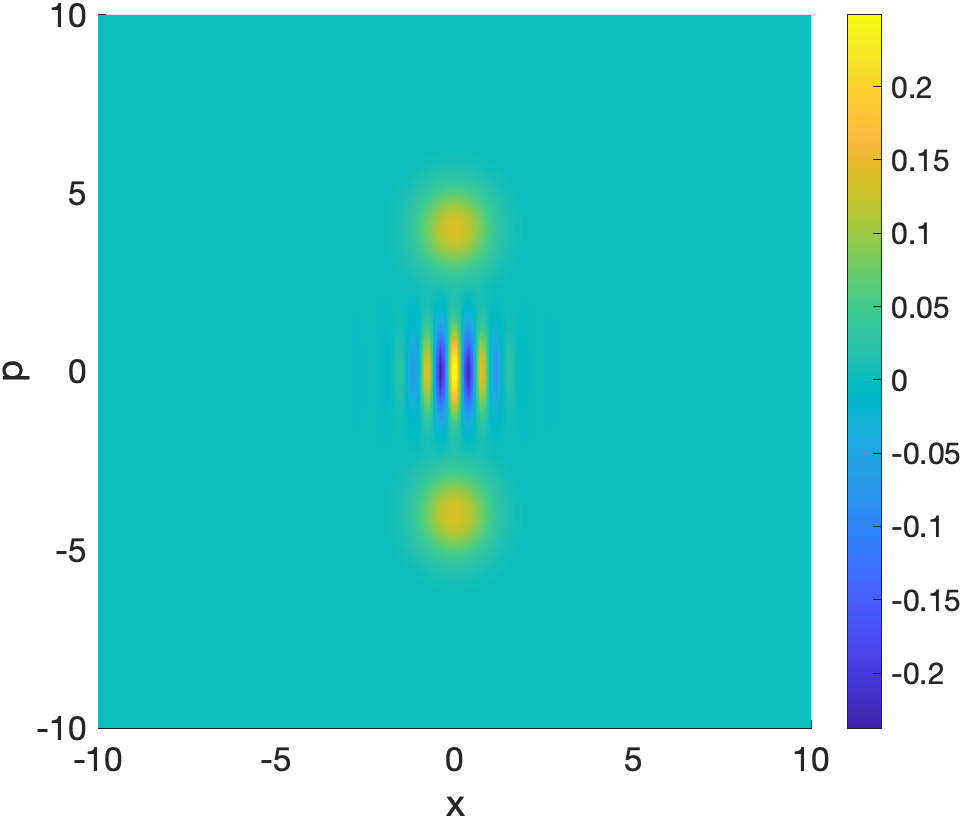}
  \hspace{-0.2cm} \includegraphics[width=0.16\textwidth]{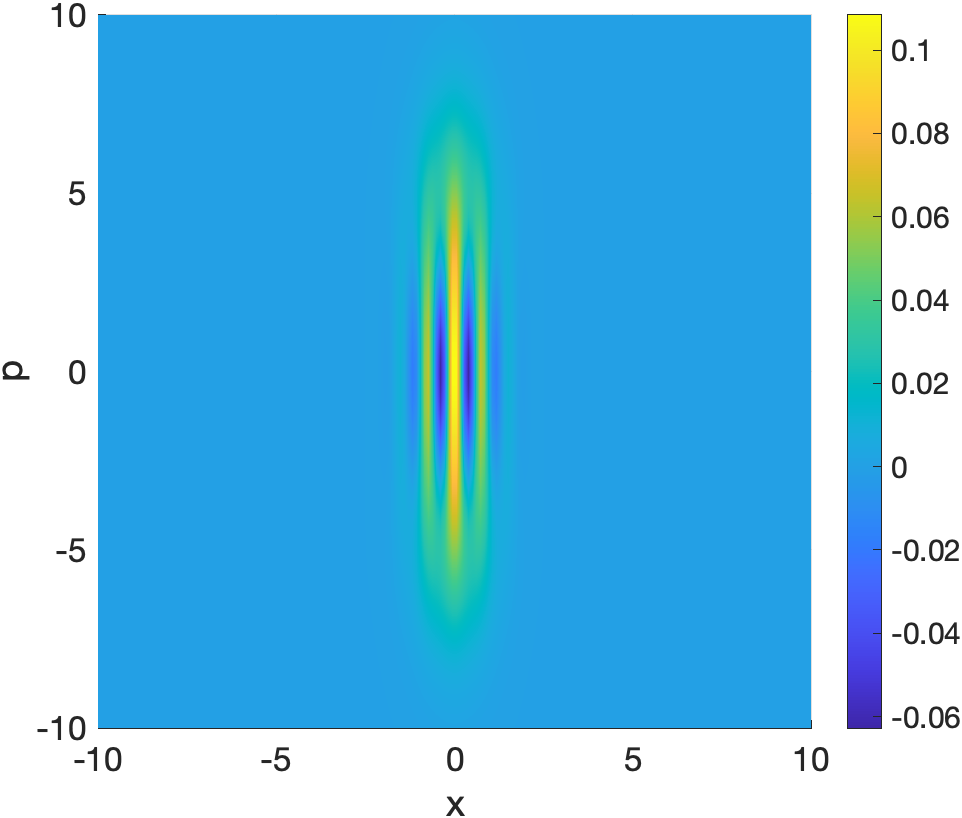}\\
  \hspace{-0.2cm} \includegraphics[width=0.16\textwidth]{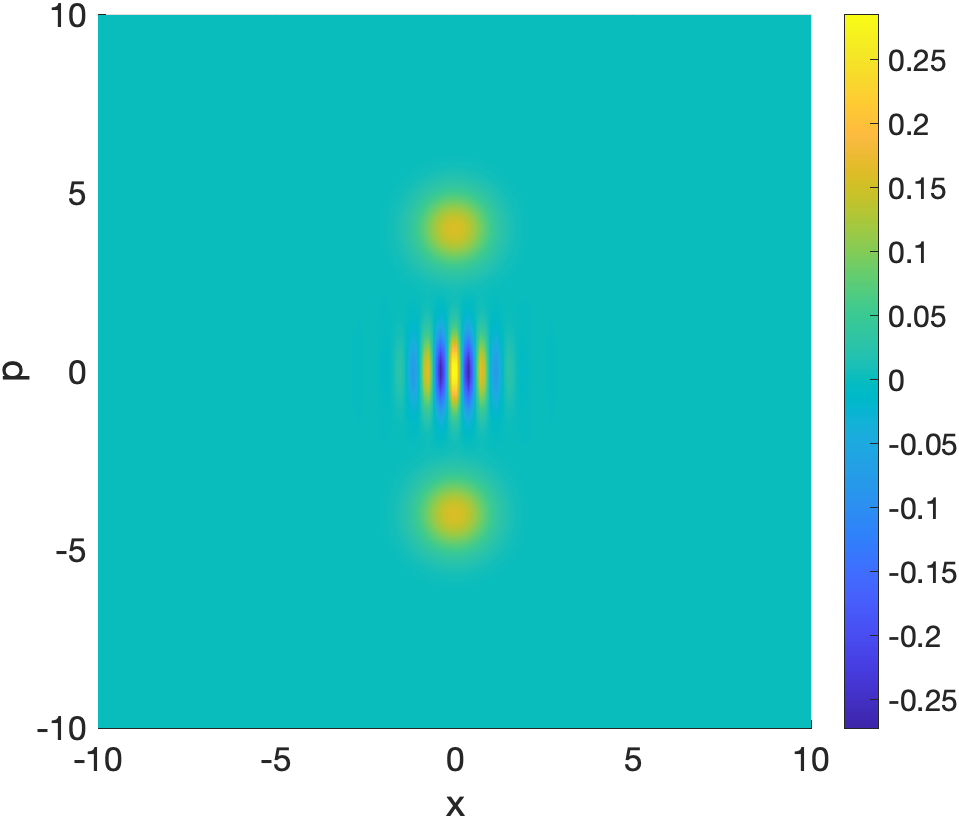} 
 \hspace{-0.2cm} \includegraphics[width=0.16\textwidth]{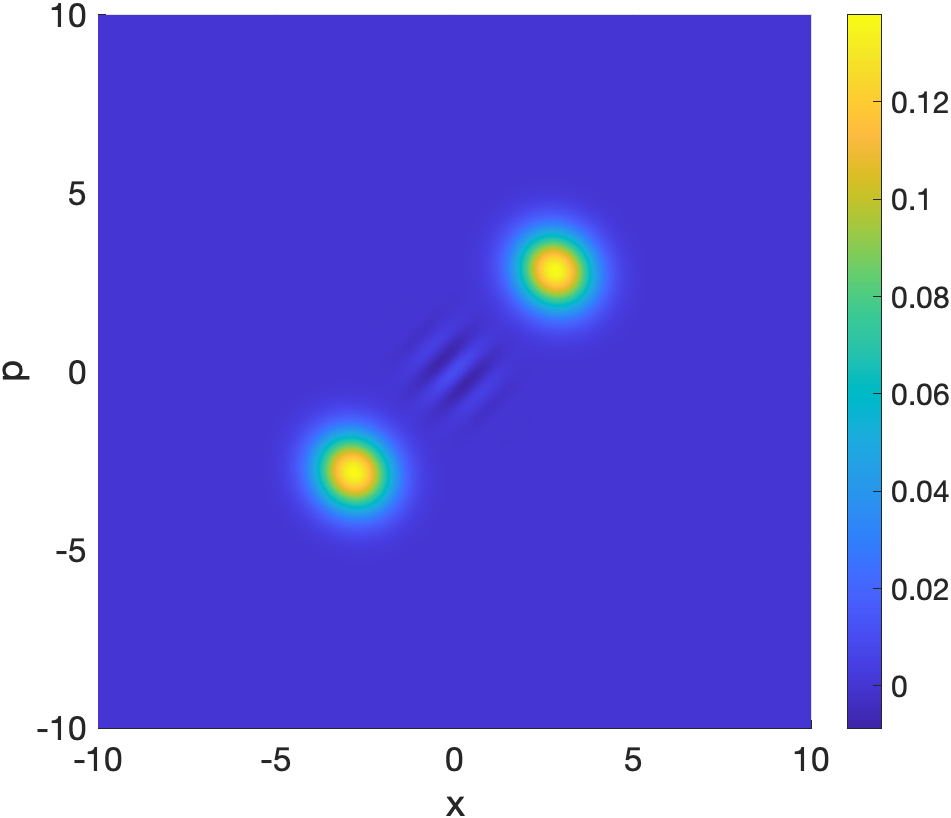}
  \hspace{-0.2cm} \includegraphics[width=0.16\textwidth]{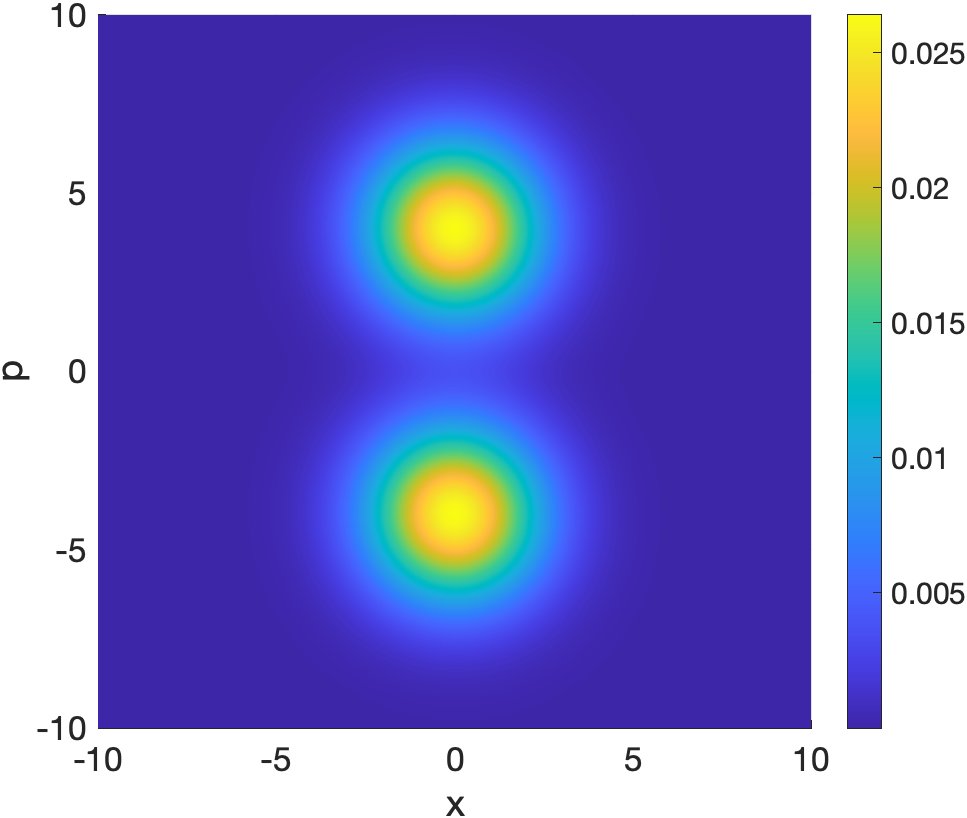}
\caption{\textit{Snapshots of the Wigner function evolved under the Lindblad evolution 
(\ref{eqn-Lindblad_q}) for $\gamma=0.2$.} 
False colour plot of the Wigner functions at times $t=0$, $t=\pi/4$, and $t=8\pi$ 
(from left to right). The top row show the time evolution of the 
Wigner function of an initial ``cat'' state in position. The middle row corresponds 
to an initial state that is a superposition of coherent states centred at two different 
momenta, and the bottom row shows the dynamics with an additional harmonic oscillator 
Hamiltonian with $\omega=1=m$ for the same initial state as in the middle row.}
\label{fig:1} 
\end{figure}

Classical physics, based on Hamiltonian mechanics, on the other hand, is modelled on phase 
space, not merely on position space. Indeed, classically, a particle moving in two different 
directions at the same time is equally unlikely as a particle being in two different positions at 
the same time. Yet, a position-space decoherence can maintain momentum 
coherence, as illustrated in Figure~\ref{fig:1} in terms of the negativity of the Wigner functions. 
Further, in the case of a cloud of quantum particles 
the exchange of momentum through scattering can also be important, and one would 
expect decoherence in both position and momentum. However, the problem is that one 
cannot simultaneously determine the position and the momentum with an arbitrary precision 
in quantum mechanics --- at best, if both were measured together \cite{Arthurs}, the 
measurement 
accuracy will be bounded by the Heisenberg relation. Paraphrasing the problem, while decoherence 
means diagonalisation of the density matrix in the basis of a preferred observable, when there 
are two incompatible preferred observables involved, it is not clear in which basis should 
the density matrix decohere. 

Here we address this issue by considering a 
coherent-state-based positive operator-valued measure (POVM) for phase-space measurements \cite{Davies,Peres}, and 
identify its effect on decoherence. Such a measurement has the property that 
outcomes are phase-space points, while the state of the system, after an outcome is 
recorded, is a coherent state centred at that point. 
Starting from an arbitrary given initial state, we show that after a single POVM 
measurement, quantum features of the initial state are washed out in the sense that 
the Wigner function as well as the associated $P$-function (cf. \cite{Vogel,Agudelo}) 
become positive. We then derive the effect of a 
repeated phase-space measurement on the density matrix, which shows that in 
a phase-space decoherence not only the off diagonal elements decay but also 
the diagonal elements are damped and converge toward a Gaussian distribution with 
ever increasing width, both in position-space and momentum-space 
representations. Thus a phase-space decoherence provides a democratic representation 
of position and momentum. 
We then show that a Lindblad equation, where position and momentum appear as 
two independent Lindblad operators, unravels the phase-space POVM measurements. 
The construction of a Lindbladian dynamical model 
that unravels a POVM measurement in itself has hitherto remained an open challenge 
(cf. \cite{Giovannetti}). Hence our results can be used as 
a basis for further investigations into developing models that unravel non-orthogonal 
measurements.  

The present paper is organised as follows. After a brief discussion on properties of 
a position-based Lindblad equation for position-space decoherence, we review the 
notion of phase-space 
measurement using POVMs. By considering the case in which a measurement is 
performed but the outcome not recorded (i.e. environmental monitoring), we are able to read off 
the impact of phase-space decoherence. We work out the effect of a repeated POVM 
measurement on an arbitrary initial density matrix, expressed in the position representation. 
By transforming to the corresponding Wigner function on phase space, we show that 
after a single POVM measurement, both the Wigner function and the $P$-function become 
positive. We then present 
a Lindblad equation that unravels the phase-space POVM measurement, and solve this in 
terms of the Wigner function. We conclude with a remark on how our approach can be 
extended to investigate phase-space decoherence for spin systems. 

Let us begin by discussing some features of decoherence induced 
by a monitoring of the position of a quantum particle by its environment.
The idea is that if the position of the system is 
measured but the outcome not recorded, then the probability distribution 
turns classical. For example, for an initial pure state 
$\psi(x)=\langle x|\psi\rangle$, 
the unrecorded monitoring of the position turns the state into a 
mixed state with density matrix $\rho(x,y)=\langle x|{\hat\rho}
|y\rangle=|\psi(x)|^2\delta(x-y)$, in which all quantum coherences are removed. 
This is the asymptotic state for the evolution generated by a Lindblad equation 
\begin{equation}
\label{eqn-Lindblad_q}
\partial_t{\hat\rho}
=\gamma\left( {\hat q}{\hat\rho}{\hat q}-\half({\hat q}^2{\hat\rho}+{\hat\rho}{\hat q}^2)\right),
\end{equation} 
where $\gamma>0$. In position representation the solution is 
${\rho}_t(x,y) = \re^{-\frac{1}{2}\gamma(x-y)^2t}{\rho}_0(x,y)$,
showing an exponential damping of the off-diagonal elements of the density matrix, 
which asymptotically approaches the fully decohered state. 
Figure \ref{fig:1} depicts three examples of cat states evolved with a position-space 
Lindblad dynamics, in Wigner representation $W(x,p)$, defined as the 
 the inverse Weyl transform of the density matrix:
\begin{eqnarray}
W(x, p) = \frac{1}{\pi}\int_{-\infty}^\infty 
\rho(x+\nu, x-\nu)\re^{-2\mathrm{i}p\nu}\; \rd\nu.
\label{eq:zz10}
\end{eqnarray} 
The Wigner function is a useful tool in the analysis of the quantum-to-classical transition, 
due to its negativity signalling quantum features. In the first column depicting 
the initial states, the typical interference patterns of cat states 
are clearly visible. The top row depicts snapshots of the time-evolved Wigner function 
under pure position-space decoherence of an initial 
cat state superposing two different positions after a relatively short time ($t=\pi/4$, middle) 
and a long time ($t=8\pi$, right), showing the familiar decay of the interference fringes. 
When the initial state is a superposition of coherent states centred at two different 
momenta (middle panels), a position-based decoherence maintains 
quantum interference. If the Lindbaldian dynamics is superposed with an underlying unitary motion that 
mixes position and momentum, such as a quantum harmonic oscillator (bottom panels), 
then owing to the mixing a position-based decoherence will eliminate decoherence in both 
directions. 

In what follows we shall investigate the generalisation of these behaviours using a decoherence 
based on phase-space measurement. For this purpose, let us 
begin by considering a normalised coherent state $|z\rangle$ defined as an eigenstate 
of the harmonic oscillator annihilation operator: ${\hat a}|z\rangle=z|z\rangle$. The coherent 
states are in one-to-one correspondence with points of classical phase space via the 
identification $z=\frac{1}{\sqrt{2}}(x+\ri p)$. (We work in scaled units with $\hbar=1$, and 
where $x$ and $p$ have the same dimension.) In terms of the harmonic 
oscillator energy eigenstates $\{|n\rangle\}$ the coherent state $|z\rangle$ admits a 
series expansion 
\begin{eqnarray}
|z\rangle=\re^{-\frac{1}{2}{\bar z}z}\sum_{n=0}^\infty \frac{z^n}{\sqrt{n!}}\, |n\rangle.
\end{eqnarray}
Coherent states form a resolution of the identity,
\begin{eqnarray}
\frac{1}{2\pi} \int |z\rangle\langle z| \, \rd^2 z ={\hat{\mathds 1}} ,
\end{eqnarray}
where $\rd^2 z = \ri \, \rd z\wedge\rd{\bar z}=\rd x\,\rd p$. Hence, they can be used to form 
a positive operator-valued measure over the classical phase space \cite{Gazeau}. 

Specifically, if a system is initially in a state represented by a density matrix 
${\hat\rho}_{\rm in}$ and 
a phase-space point $z$ is detected, then the result of the measurement is the coherent 
state $|z\rangle$ centred at that point. The probability of detecting a phase-space event 
(cf. \cite{BH0}) in the region $A$ is then given by 
\begin{eqnarray}
{\mathbb P}\left( z\in A\right) = \int_{A} {\rm tr}\left( {\hat\Pi}(z) {\hat\rho}_{\rm in} 
\right) \, \rd^2 z , 
\end{eqnarray} 
where ${\hat\Pi}(z)=(2\pi)^{-1}|z\rangle\langle z|$. Therefore, the 
``expectation'' $Q(z)= {\rm tr}( {\hat\Pi}(z) {\hat\rho}_{\rm in} )$ of the initial state 
${\hat\rho}_{\rm in}$ in the 
coherent state, known as the Husimi (pronounced Fushimi) 
density, defines a probability distribution 
over the phase space. It follows that if a phase-space measurement is performed but the 
outcome not recorded, then the state of the system decoheres into 
\begin{equation}
{\hat\rho}_{\rm out} = \int Q(z) \, {\hat\Pi}(z) \, \rd^2 z . 
\label{eq:4} 
\end{equation}
In other words, the result of a phase-space decoherence is the average of the 
coherent state projectors over the phase space with the Husimi density function $Q(z)$. 

Because the effect of decoherence is often captured in position space 
or in terms of the Wigner function \cite{Davidovich,Murakami}, 
let us transform the result (\ref{eq:4}) into these representations. 
Writing $\rho_{\rm in}(x,y) = \langle x|{\hat\rho}_{\rm in}|y\rangle$ for the 
position representation of the density matrix, we see that 
the Husimi density can be written in the form 
\begin{eqnarray}
Q(z) = \iint \rho_{\rm in}(x,y) \langle y|{\hat\Pi}(z)|x\rangle \, \rd x \, \rd y . 
\label{eq:5} 
\end{eqnarray} 
Substituting (\ref{eq:5}) in (\ref{eq:4}) and making use of the position representation 
$\langle x|z\rangle=\pi^{-\frac{1}{4}}\, \re^{-\frac{1}{2}(x-q)^2+{\rm i}p(x-q)}$ 
of the coherent state, we deduce, after a calculation involving Gaussian 
integrations, a simple formula that represents the effect of phase space decoherence 
in the form  
\begin{equation} 
\rho_{\rm out}(x,y) =  \frac{\re^{-\frac{1}{2}(x-y)^2}}{\sqrt{2\pi}} \!\! \int\limits_{-\infty}^\infty 
\re^{-\frac{1}{2}\lambda^2} \rho_{\rm in}(x+\lambda,y+\lambda) \, \rd \lambda . 
\label{eq:7} 
\end{equation}
Hence in position space there is an overall 
Gaussian damping of the off-diagonal elements of the density matrix, while 
along each diagonal, elements are averaged with a Gaussian weight. 

The effect of decoherence in (\ref{eq:7}) can be contrasted with that generated by a 
position monitoring (\ref{eqn-Lindblad_q}), 
where we have ${\rho}_t(x,y) = \re^{-\frac{1}{2}\gamma(x-y)^2t}
{\rho}_0(x,y)$. 
Over a short time (say, $t=1$), there is an identical Gaussian damping of the 
off-diagonal elements as in (\ref{eq:7}), whereas the Gaussian smoothing along 
the diagonals in (\ref{eq:7}) is a feature resulting from phase-space monitoring. 
Importantly, as expected the effect of a phase-space decoherence is identical in the 
position and the momentum representation, which can be easily seen 
for example by taking a double Fourier transform of (\ref{eq:7}). 
The same cannot be said for the coordinate-space decoherence 
modelled by the solution to 
the position-based Lindblad equation. 

As a matter of interpretation, we can regard the state (\ref{eq:4}), or equivalently 
(\ref{eq:7}), as representing 
classicality that is more robust against environmental perturbation (cf. \cite{BH}), 
with a higher entropy than the initial state such that the expectation values of both 
the position and the momentum are identical to their initial values. On the other hand, 
owing to the lack of orthogonality in POVM measurements, ${\hat\rho}_{\rm out}$ 
does not fulfil Zurek's repeatability criterion \cite{Zurek2018}, that repeated 
measurements give rise to the same outcome as the initial measurement. In 
practical terms, what this 
means is that the outcome state ${\hat\rho}_{\rm out}$ of (\ref{eq:4}), or equivalently 
its position representation (\ref{eq:7}), need not be the terminal state of phase 
space decoherence: We can substitute $\rho_{\rm out}(x,y)$ of (\ref{eq:7}) into 
$\rho_{\rm in}(x,y)$ of its integrand, and repeat this procedure. Then after $m$ 
iterations we find that 
\begin{equation} 
\rho^{(m)}(x,y) =  \frac{\re^{-\frac{m}{2}(x-y)^2}}{\sqrt{2\pi m}} \!\!
\int\limits_{-\infty}^\infty 
\re^{-\frac{\lambda^2}{2m}} \rho_{\rm in}(x+\lambda,y+\lambda) \, \rd \lambda . 
\label{eq:8} 
\end{equation}
Hence as the process of decoherence is repeated, in position space the 
off-diagonal elements of the density matrix are suppressed exponentially, at the rate 
given by the square of the distance $|x-y|$ away from the diagonal. Along each 
diagonal, the matrix elements are averaged with respect to a Gaussian density 
with the standard deviation increasing in $m$. In particular, 
if we let $X$ denote a random variable with the density $\rho_{\rm in}(x,x)$, and 
let $\{N_j\}$ be a set of independent and identically distributed Gaussian random 
variables with mean zero and variance one, then $\rho^{(m)}(x,x)$ represents the 
density function of the random variable $X+\sum_{j=1}^m N_j$. 
The entropy of the system will thus increase along the 
way without bound. 

To investigate the effect of decoherence in phase space, we consider again the 
Winger quasi-probability distribution over the phase space.  
(An alternative formulation is to employ the 
Bargmann-Husimi transform \cite{Bargmann,Husimi}, which will be develop elsewhere.) 
To this end we first 
express the initial density matrix in terms of the corresponding Wigner function 
using the Weyl transform
\begin{eqnarray}
\rho_{{\rm in}}(x,y) = \int_{-\infty}^\infty W_{{\rm in}} \left(\frac{x+y}{2}, \mu\right)
\re^{\mathrm{i}\mu(x-y)}\; \rd \mu . 
\label{eq:9}
\end{eqnarray} 
We substitute (\ref{eq:9}) in (\ref{eq:8}) to express 
$\rho^{(m)}_{{\rm out}}$ in terms of the initial Wigner function, and then substitute 
the result in (\ref{eq:zz10}). After rearrangement of terms and performing one of 
the integrations, we deduce the effect of phase-space decoherence on the Wigner 
function due to $m$ successive POVM measurements as  
\begin{equation}
W_{\textrm{out}}^{(m)}(x, p) = \frac{1}{2\pi m} \iint\limits_{-\infty}^{~~~\infty} 
\re^{\frac{-(p-k)^{2}}{2m}} \re^{-\frac{(x-q)^{2}}{2m}}W_{{\rm in}}(q, k) 
\, \rd q \rd k . 
\label{eq:11}
\end{equation} 

By setting $m=1/2$, substituting (\ref{eq:9}) in (\ref{eq:5}), and performing a short 
calculation, it becomes evident that the right side of (\ref{eq:11}) is nothing but 
the Husimi density function (\ref{eq:5}) of the initial density matrix. 
Because a Husimi density is always positive, it follows that quantum 
features of the initial state as represented by the negativity of the Wigner function is already 
eliminated by a half iteration. The positivity of the Wigner function, of course, is not a 
sufficient condition for classicality. For the latter, it is more common to require the 
Glauber-Sudarshan $P$-function to become positive \cite{Vogel,Agudelo}. 
Because the Wigner function is the double-Gaussian 
convolution of the $P$-function of the form (\ref{eq:11}) with $m=1/2$ \cite{Diosi}, it 
follows that the $P$-function of $W^{(1)}$ is just $W^{(1/2)}$, which is positive. It follows that 
after a single POVM measurement on phase space, the $P$-function becomes positive, 
thus quantum features have been eliminated. 

Our next task is to propose a dynamical model that gives rise to phase-space 
decoherence. We show that the Lindblad equation with two Lindblad operators, 
one the position ${\hat L}_1=\sqrt{\gamma}{\hat x}$, and the other the momentum operator 
${\hat L}_2=\sqrt{\gamma}{\hat p}$, gives rise to the right dynamics for our purpose, in line 
with the interpretation that a Hermitian Lindbladian corresponds to a monitoring of the 
relevant observable. Thus consider a purely dissipative Lindblad evolution for the 
density matrix of the form (setting $\gamma=1$ to avoid clutter) 
\begin{equation}
\frac{\partial{\hat\rho}}{\partial t} = {\hat x} {\hat\rho}{\hat x} - \half 
\left({\hat x}^2{\hat\rho}+{\hat\rho}{\hat x}^2\right)+ {\hat p} 
{\hat\rho}{\hat p}- \half \left({\hat p}^2{\hat\rho}+{\hat\rho}{\hat p}^2\right).
\label{eq:12} 
\end{equation}
Let us solve this equation by transforming to the Wigner representation, in which the 
equation becomes 
\begin{eqnarray}
\frac{\partial W}{\partial t} &=& x\star W\star x- \half \left(x\star x\star W+ W\star x\star x 
\right) \nonumber \\ && + p\star W\star p- \half\left(p\star p\star W+ W\star p\star p\right), 
\,\, 
\label{eq:x13}
\end{eqnarray}
where $\star$ denotes the Moyal product \cite{Curtright}. Due to the linearity of 
${\hat L}_1$, ${\hat L}_2$ in ${\hat p}$ and ${\hat x}$, (\ref{eq:x13}) simplifies to
\begin{eqnarray}
\frac{\partial W}{\partial t} = \frac{1}{2}
\frac{\partial^2 W}{\partial p^2}+\frac{1}{2}
\frac{\partial^2 W}{\partial x^2} . 
\label{eq:15}
\end{eqnarray}
This can be solved by transforming to the Fourier space. Specifically, writing 
\begin{equation}
\chi(q,k)=\frac{1}{2\pi}\iint {\re}^{-{\rm i}qx}{\re}^{-{\rm i}kp}\, W(x,p)\, {\rd}x{\rd}p ,
\end{equation}
(\ref{eq:15}) transforms into 
\begin{equation}
\frac{\partial}{\partial t}\chi_t(q,k) = -\frac{1}{2}\left(q^2+k^2\right)\chi_t(q,k) ,
\end{equation}
the solution of which is given by 
\begin{equation}
\chi_t(q,k)={\re}^{-\frac{1}{2}(k^2+q^2)t} \, \chi_0(q,k).
\end{equation}
Performing an inverse Fourier transform, we thus find the solution to (\ref{eq:15}) 
as 
\begin{equation}
W_{t}(x, p) = \frac{1}{2\pi t} \iint\limits_{-\infty}^{~~~\infty} 
\re^{\frac{-(p-k)^{2}}{2t}} \re^{-\frac{(x-q)^{2}}{2t}}W_{0}(q, k) \, \rd q \rd k .
\label{eq:19}
\end{equation} 
For integer values of $t=m$ the solution indeed corresponds to (\ref{eq:11}). 

\begin{figure}[t]
      \centering
\hspace{-0.2cm} \includegraphics[width=0.16\textwidth]{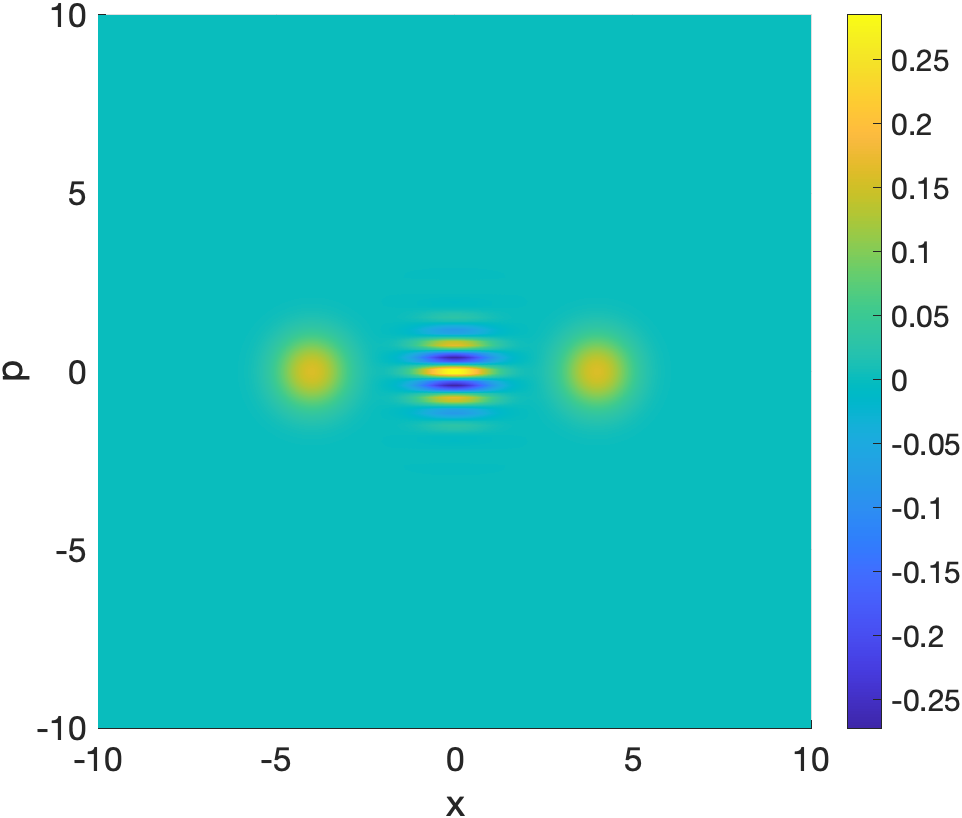} 
 \hspace{-0.2cm} \includegraphics[width=0.16\textwidth]{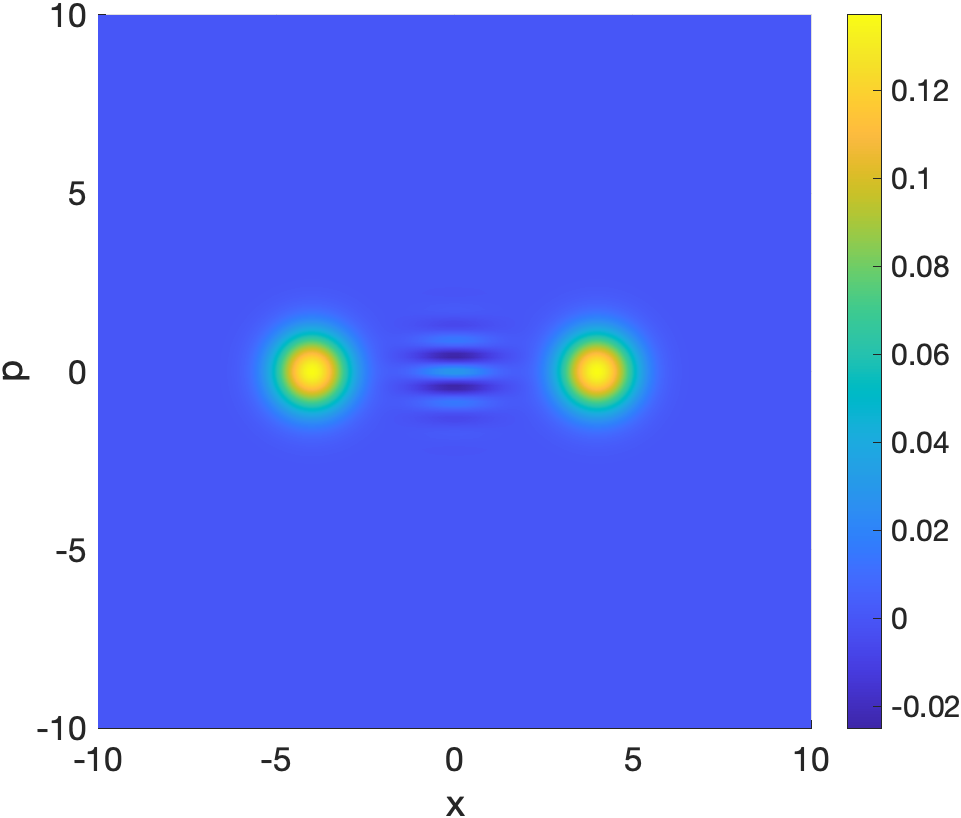}
  \hspace{-0.2cm} \includegraphics[width=0.16\textwidth]{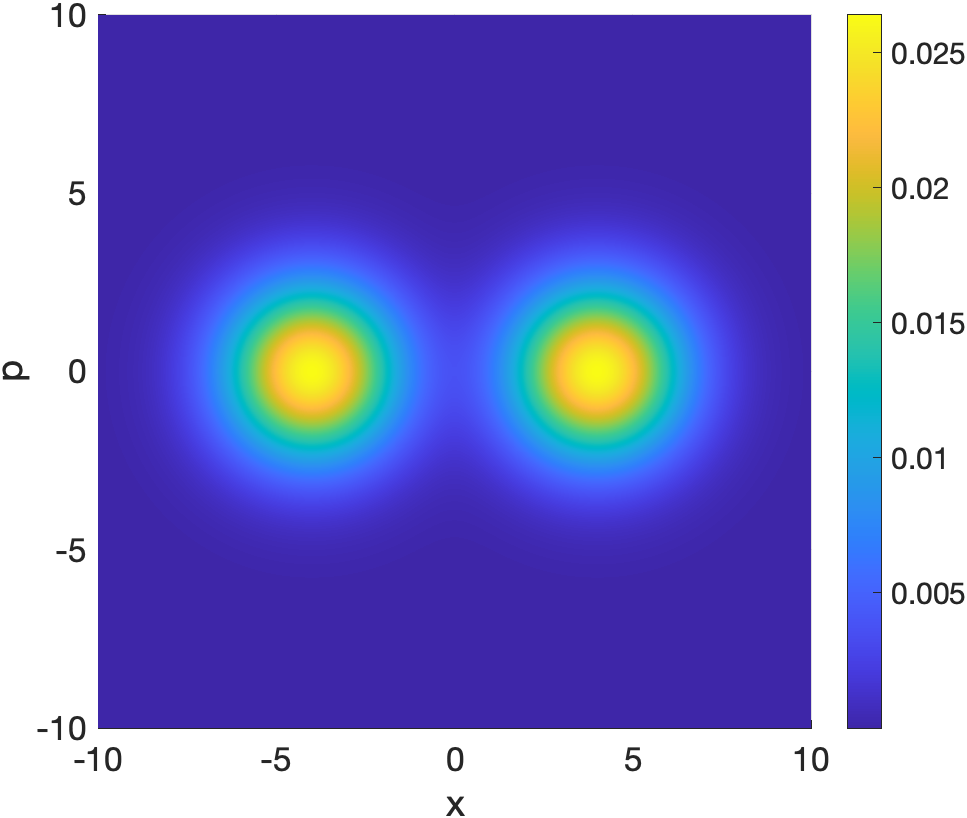}\\
  \hspace{-0.2cm} \includegraphics[width=0.16\textwidth]{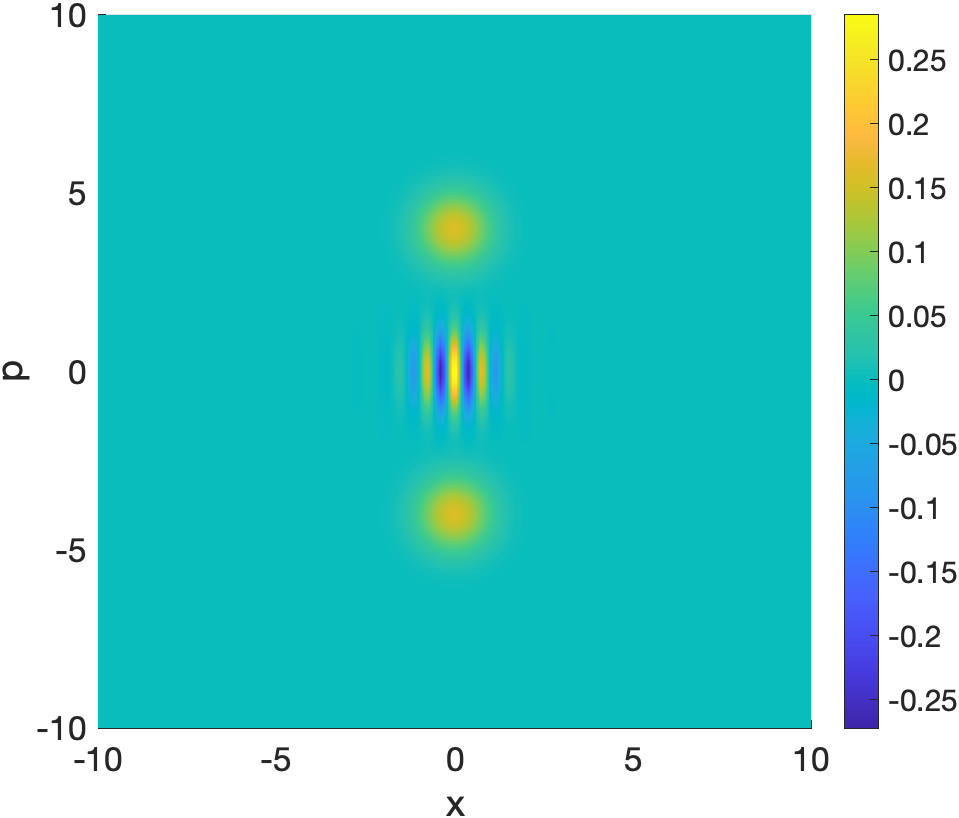} 
 \hspace{-0.2cm} \includegraphics[width=0.16\textwidth]{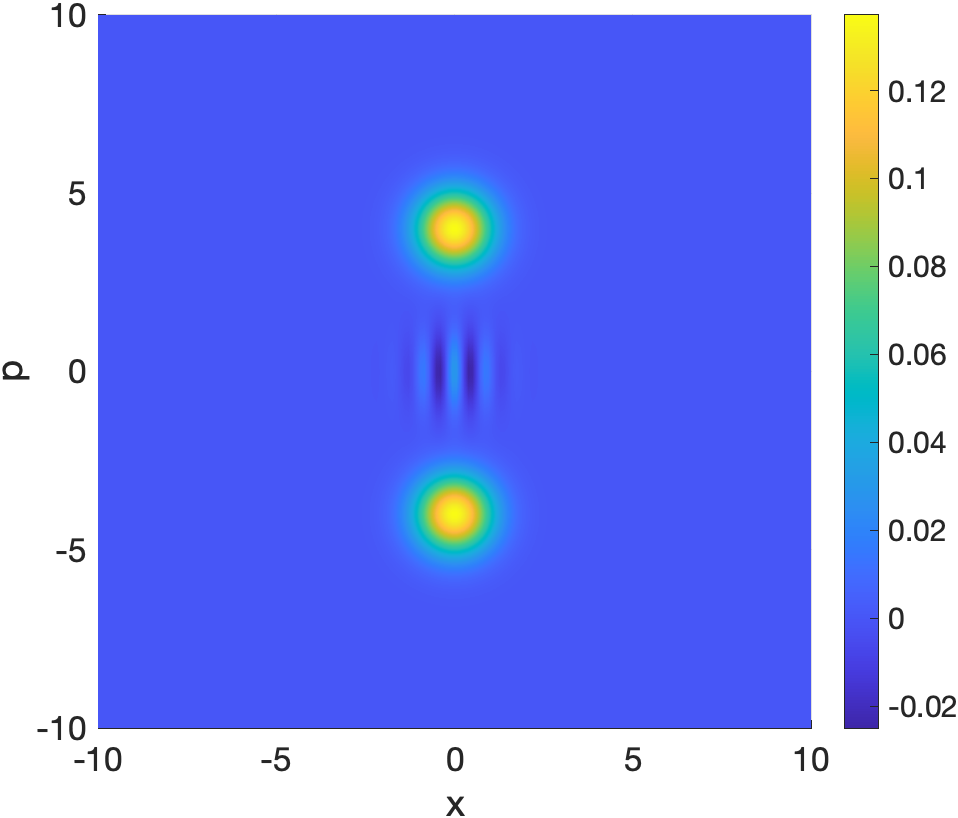}
  \hspace{-0.2cm} \includegraphics[width=0.16\textwidth]{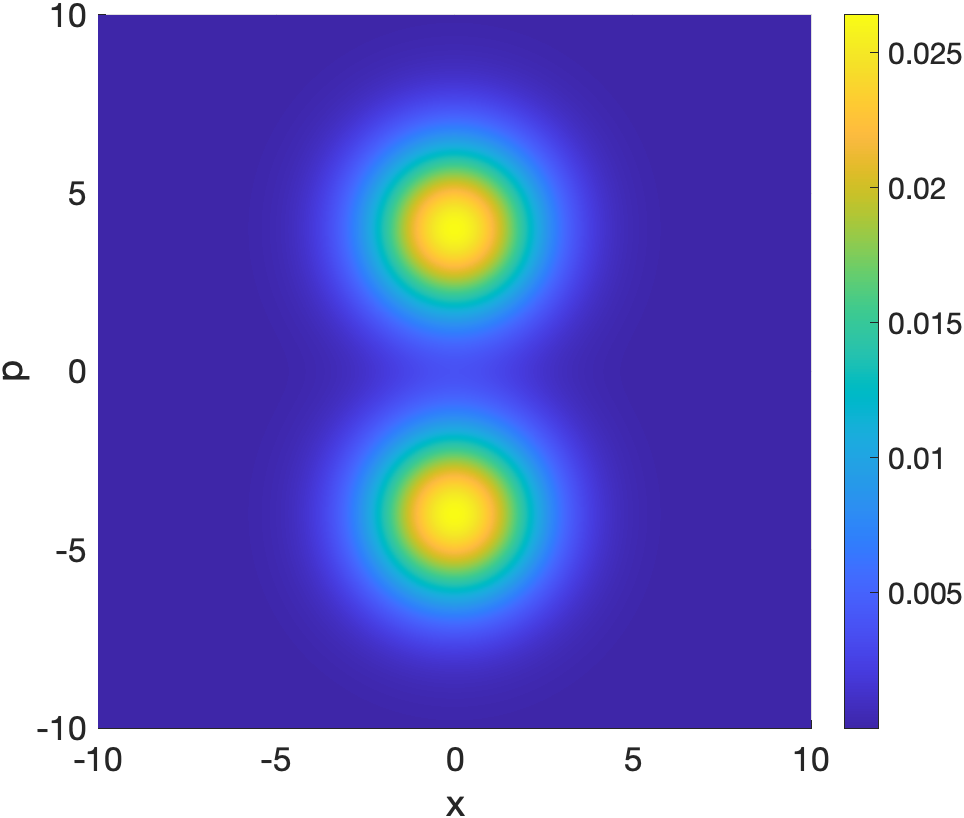}\\
  \hspace{-0.2cm} \includegraphics[width=0.16\textwidth]{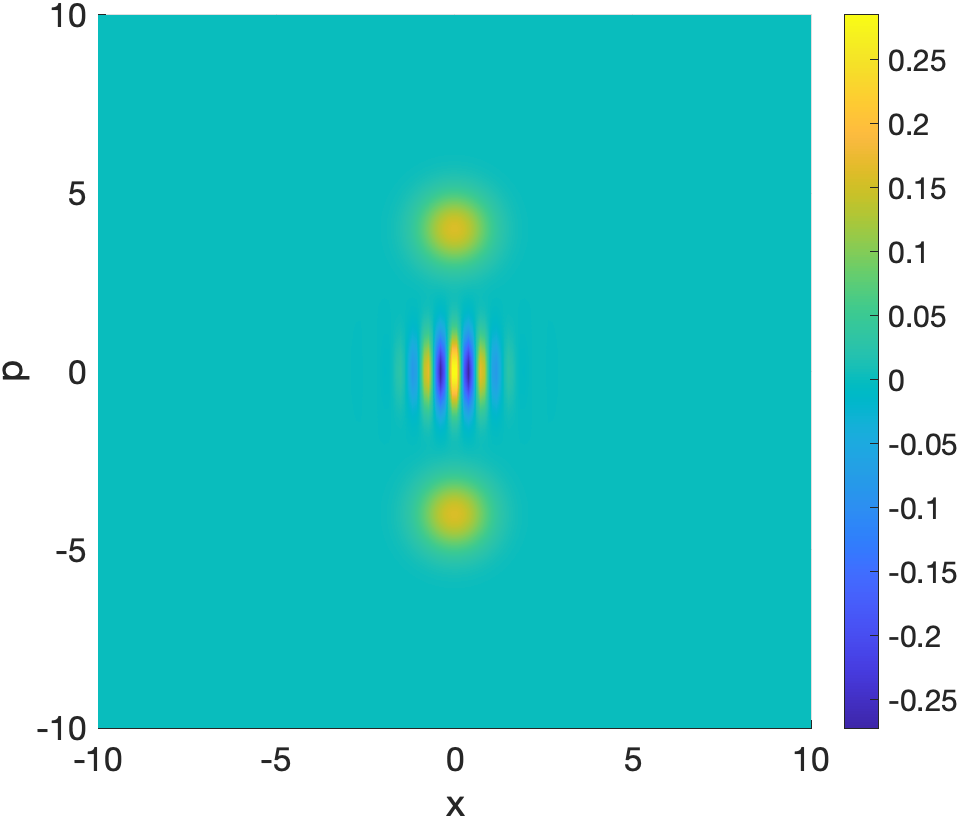} 
 \hspace{-0.2cm} \includegraphics[width=0.16\textwidth]{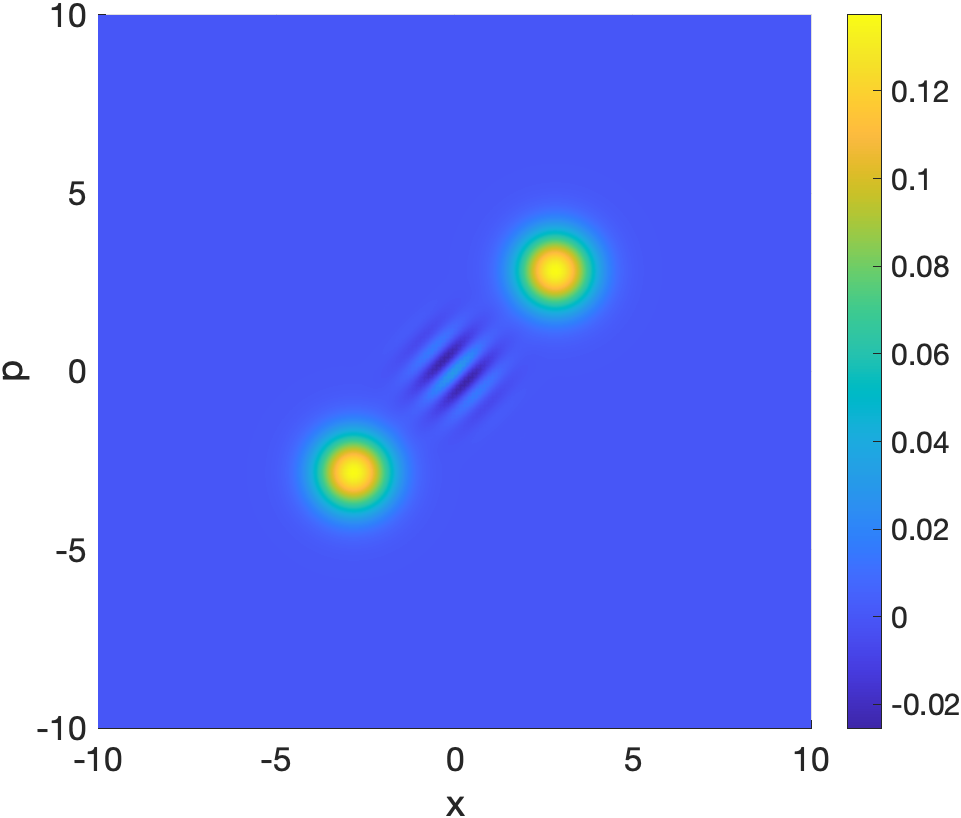}
  \hspace{-0.2cm} \includegraphics[width=0.16\textwidth]{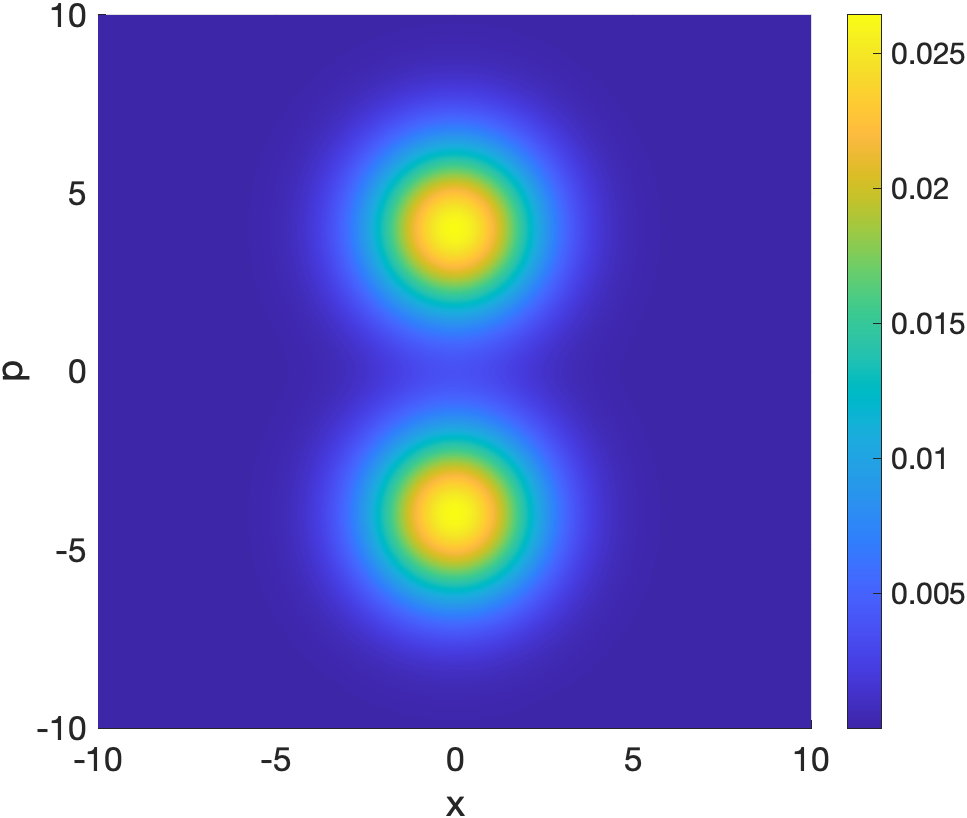}
\caption{\textit{Density plot of the Wigner function under the Lindblad evolution for 
phase-space decoherence}. All parameters and initial conditions are the same as 
in Figure~\ref{fig:1}, except for the dynamical evolution generated by the Lindblad 
equation (\ref{eq:12}), with $\gamma=0.1$.}
\label{fig:2} 
\end{figure}

In Figure~\ref{fig:2} we show the dynamical behaviour of the Wigner function associated 
with solutions to the Lindblad equation for phase-space decoherence, to be contrasted 
with Figure~\ref{fig:1}. For an arbitrary initial state, interference fringes are 
eliminated across the whole of the phase space.  The top panels show the time evolution 
(from left to right) of the Wigner function of an initial ``cat'' state for two different positions. 
For a superposition of coherent states centred at two different 
momenta (middle panels), a phase-space decoherence eliminates interference. 
When the Lindbaldian dynamics is superposed with an underlying unitary motion 
generated by a harmonic oscillator Hamiltonian (bottom panels), the 
behaviour of the position-space decoherence in Figure~\ref{fig:1} 
is very similar to that of a full phase-space decoherence in Figure~\ref{fig:2}. 
This can be understood by considering the Lindblad equation 
in a moving frame where the harmonic oscillator motion is translated into a time-dependence 
in the Lindbladians, which leaves the phase-space decoherence Lindblad equation invariant, 
but which dynamically oscillates between position and momentum decoherence in the 
position-space decoherence model.

So far we have taken a democratic approach in position and 
momentum variables. We can, however, reinstate some 
of the physical parameters by introducing a variable $\sigma=\sqrt{\hbar/\omega}$ in terms 
of the Planck constant and the frequency of the harmonic oscillator to redefine our 
coherent state $|z\rangle$ leading to a stronger decoherence effect in either position or momentum. 

In summary, we have constructed a scheme for modelling decoherence in phase space 
by use of a coherent-state-based POVM. Our model shows that under phase-space 
decoherence the density matrix of the system is diagonalised in both the position and 
the momentum representations, such that the diagonal elements are also smoothed 
with respect to a standard Gaussian measure. The expectation values of both position 
and momentum operators are conserved during the decoherence process. 
We have presented an explicit dynamical model for phase space decoherence in the 
form of a simple Lindblad equation (\ref{eq:12}) that is analytically solvable. 
The solution in terms of the Wigner function shows that starting 
from an arbitrary initial state, after a single POVM measurement the Wigner function and 
even the associated P-function of the resulting state become strictly positive. 
The construction of 
a Lindbladian model for a POVM measurement in itself has hitherto remained an open 
question, to which our result may provide useful insights. 

We conclude by remarking that the method employed here will prove useful in 
other studies on the emergence of classicality, for example, for spin systems 
for which the corresponding coherent states are well understood 
\cite{Peremolov1972}. Indeed, for spin systems we can model phase-space decoherence 
by constructing POVM measurements using the ${\rm SU(2)}$ coherent states \cite{BH2}. 
It appears that the effect of such decoherence can be unravelled dynamically using a Lindblad 
equation generated by the three angular momentum operators ${\hat S}_x$, 
${\hat S}_y$, and ${\hat S}_z$ --- details of which will be developed elsewhere.

\vspace{0.2cm} 
\noindent 
DCB acknowledges support from EPSRC (grant EP/X019926/1) and the John 
Templeton Foundation (grant 62210). The opinions expressed in this 
publication are those of the authors and do not necessarily reflect the 
views of the John Templeton Foundation. EMG acknowledges support 
from the Royal Society (Grant. No. URF/R/201034). 
RM is funded through an Imperial College President’s PhD Scholarship.


\begin{thebibliography}{}

\bibitem{Zurek1991}
W.~H.~Zurek,  
Decoherence and the transition from quantum to classical, 
{\em Phys. Today} \textbf{44} (10), 36 (1991). 

\bibitem{Zurek} 
Zurek,~W.~H. (2002) 
Decoherence and the Transition from Quantum to Classical--Revisited. 
Los Alamos Science \textbf{27}. 

\bibitem{JZ} 
E.~Joos, H.~D.~Zeh, C.~Kiefer, D.~Giulini, J.~Kupsch \& I.~O.~Stamatescu,  
{\em Decoherence and the Appearance of a Classical World in Quantum Theory}, 
2nd ed. (Springer, Berlin, 2003). 

\bibitem{MS} 
M.~Schlosshauer,
{\em Decoherence and the Quantum-to-Classical Transition} (Springer, Berlin, 2010). 

\bibitem{Zurek22}
W.~H.~Zurek,
Quantum theory of the classical: einselection, envariance, quantum Darwinism and 
extantons, 
{\em Entropy} \textbf{24}, 1520 (2022).

\bibitem{Halliwell} 
Di\'osi,~L. \& Halliwell,~J.~J. (1998) 
Coupling classical and quantum variables using continuous quantum measurement 
theory. 
{\em Physical Review Letters} \textbf{81}, 12846-2849.

\bibitem{Schlosshauer 2019} 
M.~Schlosshauer,   
Quantum decoherence, 
{\em Physics Reports} \textbf{831}, 1-57 (2019). 

\bibitem{Arthurs} 
Arthurs,~E. \& Kelly,~Jr.,~J.~L. (1965) 
Simultaneous measurement of a pair of conjugate observables. 
{\em The Bell System Technical Journal} \textbf{44}, 725-729. 

\bibitem{Davies} 
Davies,~E.~B. (1976) 
\textit{Quantum Theory of Open Systems}. 
(London: Academic Press). 

\bibitem{Peres} 
Peres,~A. (2006) 
\textit{Quantum Theory: Concepts and Methods}. 
(New York: Kluwer Academic Publishers).

\bibitem{Vogel} 
Ryl,~S., Sperling,~J., Agudelo,~E., Mraz,~M., K\"ohnke,~S., Hage,~B., \& Vogel,~W. 
(2015) 
Unified nonclassicality criteria. 
{\em Physical Review} A\textbf{92}, 011801. 

\bibitem{Agudelo} 
Bohmann,~M., Agudelo,~E., \& Sperling,~J. (2020) 
Probing nonclassicality with matrices of phase-space distributions. 
{\em Quantum} \textbf{4}, 343.   
  

\bibitem{Giovannetti} 
De~Pasquale,~A., Foti,~C., Cuccoli,~A., Giovannetti,~V. \& Verrucchi,~P. (2019) 
Dynamical model for positive-operator-valued measures. 
{\em Physical Review} A\textbf{100}, 012130. 

\bibitem{Gazeau} 
Ali,~S.~T., Antoine,~J.-P. \& Gazeau,~J.-P. (2013)
{\em Coherent States, Wavelets, and Their Generalizations}. 
(Berlin: Springer). 

\bibitem{BH0} 
Brody,~D.~C. \& Hughston,~L.~P. (2021) 
Quantum measurement of space-time events. 
{\it Journal of Physics} A\textbf{54}, 235304.

\bibitem{Davidovich} 
Davidovich,~L. (1999) 
Decoherence, Wigner functions, and the classical limit of quantum mechanics 
in cavity QED. 
{\em AIP Conference Proceedings} \textbf{461}, 151-162.

\bibitem{Murakami} 
Murakami,~M., Ford,~ G.~W. \& O'Connell,~R.~F. (2003) 
Decoherence in phase space. 
{\em Laser Physics}, \textbf{13}, 180-183.  

\bibitem{BH} 
Brody, D.C. \& Hughston, L.P. (2000) 
Classical fields as statistical states. 
{\it Twistor Newsletter} \textbf{45}, 40-43.

\bibitem{Zurek2018} 
Zurek,~W.~H. (2018) 
Quantum theory of the classical: quantum jumps, Born’s Rule and objective classical 
reality via quantum Darwinism. 
{\em Philosophical Transactions of the Royal Society}, A\textbf{376}, 20180107. 

\bibitem{Bargmann} 
Bargmann,~V. (1961) 
On a Hilbert space of analytic functions and an associated integral 
transform. 
{\em Communications in Pure and Applied Mathematics} \textbf{14}, 187-214.

\bibitem{Husimi} 
Husimi,~K. (1940) 
Some formal properties of the density matrix. 
{\em Proceedings of the Physico-Mathematical Society of Japan} 
\textbf{22}, 264-314. 

\bibitem{Diosi} 
Di\'osi,~L. \& Kiefer,~C. (2002) 
Exact positivity of the Wigner and P-functions of a Markovian open system. 
{\em Journal of Physics} A\textbf{35}, 2675-2683. 

\bibitem{Curtright} 
T.~L.~Curtright \& C.~K.~Zachos (2012) 
Quantum mechanics in phase space. 
{\em Asia Pacific Physics Newsletter} \textbf{1}, 37-46. 

\bibitem{Peremolov1972} 
A.~M.~Peremolov (1972) 
Coherent states for arbitrary Lie group. 
{\em Communications in Mathematical Physics} \textbf{26}, 222-236. 

\bibitem{BH2} 
Brody,~D.~C. \& Hughston,~L.~P. (2015) 
Universal quantum measurements. 
{\em Journal of Physics: Conference Series} \textbf{624}, 012002.

\end{thebibliography}
\end{document}